\documentstyle[preprint,aps,eqsecnum]{revtex}

\begin{document}
\title{Non-linear excitations in 1D correlated insulators}
\author{Chun-Min Chang$^1$, A.~H.~Castro Neto$^{1,2}$ and A.~R.~Bishop$^3$}
\address
{$^1$ Department of Physics,
University of California,
Riverside, CA 92521 \\
$^2$ Department of Physics, Boston University, Boston, MA 02215 \\
$^3$Theoretical Division and Center for Nonlinear Studies, Los
Alamos National Laboratory, Los Alamos, New Mexico 87545}
\date{\today }
\maketitle

\begin{abstract}
In this work we investigate charge transport in one-dimensional (1D) 
insulators via semi-classical and perturbative renormalization
group (RG) methods. 
We consider the problem of electron-electron, electron-phonon
and electron-two-level system interactions. We show that non-linear  
collective modes such as polarons and solitons are reponsible for transport.
We find a new excitation in the Mott insulator: the polaronic soliton. 
We discuss the differences between band and Mott insulators in
terms of their spin spectrum and obtain the charge and spin gaps
in each one of these systems. We show that electron-electron
interactions provide strong renormalizations of the energy scales
in the problem.
\end{abstract}

\newpage

\section{Introduction}
\label{introduction}

It is with joy that we celebrate Michael Pollak's birthday. One of
us (A.H.C.N.) had the honor of inheriting Mike's office after his
retirement. In this office we had long and exciting conversations 
about many aspects of condensed matter physics and especially one
of Mike's early interests: DNA electronics. It turns out that Mike
worked for a few years as an experimentalist and he was probably
one of the first researchers to look into the problem of DNA stability
with condensed matter experimental tools \cite{pollak1}. These
were probably one of the first systematic studies of DNA molecules
from the condensed matter point of view. In fact, Mike went ahead
and also worked on some theoretical aspects of DNA
\cite{pollak2}. Besides being a complete scientist, theorist and 
experimentalist, Mike's interest in DNA was far ahead of his own time. 
Mike's broad vision of science can also be seen in his beautiful and
important work on the insulator-metal transition \cite{pollak3}. 
Perhaps because of his knowledge and experience, Mike was always convinced
that DNA was not a simple band insulator but a more exotic
Mott insulator. 
At this point in time it is not clear if a DNA molecule is an ordinary
band insulator or the more complex Mott insulator. 
We believe that the evidences that DNA is an insulator are quite strong:
DNA is transparent to light and transport measurements 
have shown a charge gap of the order of a few volts \cite{gruner,dekker}. 
From the theoretical point of view the question that remains is: 
is DNA a band insulator or a Mott insulator? 
While naive electron counting and LDA calculations \cite{lda}
seem to indicate that DNA is a band insulator with a very narrow
bandwidth, there are some experimental indications that DNA is not 
an ordinary band insulator. 
The first comes from the fact that antiferromagnetic
excitations have been observed in DNA \cite{barton} 
(see our discussion of magnetic fluctuations in Mott insulators below) 
and DNA seems to
be able to conduct supercurrents when connected to superconducting
electrodes via a proximity effect \cite{proximitydna}. 
These effects cannot be observed in ordinary insulators where the
gap to charge excitations produces a vanishing density of
states at the chemical potential. So, maybe, once again Mike was right
and DNA is hiding a few surprises.
However, the field of DNA electronics
is still in its infancy. Many more experiments are needed in
order to understand the real nature of electron propagation in DNA
and the theoretical effort is intense \cite{polarondna,fri,rudnick,cox}.
It is with an eye to these possibilities that in the last few years
we have been trying to understand the problem of conduction in 1D
insulators, which is the main theme of this paper. We are thankful to
Mike for pointing the way.

Band insulators cannot transport charge at low temperatures 
because of a charge gap in the excitation spectrum. When 
doped with acceptors and/or donors impurities, a band insulator
can conduct charge as in the case of semiconductors. But the
periodicity of the lattice is broken and impurity bands can only
be formed at high density of dopants. There is, therefore, a
competition between the disorder (which localizes the carriers),
and the overlap between the impurities wave-function (which
delocalizes the carriers). In band insulators electron-electron
interactions are supposed to be weak and are not usually discussed.
While in three dimensions this is certainly true, 1D
systems are clearly strongly interacting because electrons cannot avoid
each other in their motion, in other words, interactions are enhanced
in 1D because of phase space constraints.
In this paper we are going to discuss the
case of clean insulators in one dimension that can only transport
charge via non-linear excitations. By their nature these
non-linear excitations only exist if the electronic system interacts
with itself or 
with another set of degrees of freedom that can provide feedback effects
and hence nonlinearity.

The classical example of a non-linear excitation in a insulator is
a polaron: the dressing of a single electron in the conduction band
by phonons. Because of non-linear effects the polaron problem is
equivalent to the self-trapping of the electron and the creation of
a bound state below the conduction band (very much like a donor state
in a semiconductor). The polaron problem has a long history and it is
a well-known non-linear problem \cite{polaron}. Another famous problem is the
propagation of solitons in polymers like polyacetylene. While
polyacetylene is an insulator because of the Peierls mechanism that
leads to the doubling of the unit cell, doping can produce 
lattice-soliton
states in the middle of the gap that in principle can carry charge
\cite{Heeger,alan}. As in the case of band insulators, the Peierls
mechanism does not require any electron-electron interactions and
usually the lattice distortion is assumed to be static so polaronic
band effects are not usually discussed. In systems where insulating behavior
is driven by electron-electron interactions, that is in the case of
the so-called Mott insulators, charge-solitons are known to exist 
\cite{Luther,Heidenreich}. Usually charge-solitons are again only discussed
in the context of a static lattice without any polaronic effects.
These are collective excitations associated 
with the electronic charge density and, like polarons and lattice-solitons,
they are energetically costly since these are topological defects of
the field theory that describes them in the first place.
As we will show there are many similarities between polaron
and soliton conduction and in general these two types of excitations 
play an important role in 1D systems.

In this paper we generalize the problem of soliton formation in 1D
insulators by taking into account not only the electron-electron
interaction (which is unavoidable in these systems) but also the
electron-phonon interaction and the interaction between electrons
and two-level systems. Our interest in discussing the interaction
between two-level systems and electrons arises from three main sources.
The first is the electron interaction with hydrogen bonds in biological
systems such as DNA \cite{hbond} ,
the second is the interaction between electrons
and dangling bonds of atoms in 1D nano-structures \cite{nano}, and
finally two level systems appear in problems of magnetic moments
in 1D metals - the so-called Kondo chain \cite{kondo}. 
In all cases the electronic scattering by localized two-level systems
can lead to new effects such as unitary scattering and Kondo resonance.
Thus, two-level systems behave like pseudo-spins.
As is expected, the level of complexity of the systems we are going
to discuss is quite high but, as we are going to show, the interplay
between the electrons with these degrees of freedom leads to new effects
that cannot be observed in their absence. As we are going to
see the energy for creation of non-linear excitations is greatly
reduced by feedback effects \cite{Chang} and this might be at the
core of the problem of electron propagation in DNA molecules 
\cite{bartonzewail}.

The paper is organized as follows: in Section \ref{model} we 
describe the model; in Section \ref{band} we describe the
problem of a 1D non-interacting band insulator (this is a somewhat
idealized situation but will illustrate very well the main concepts
in polaron physics and how they apply in one-dimension);
in Section \ref{mott} we discuss the problem of non-linear transport
in a Mott insulator and compare with its non-interacting counterpart;
in Section \ref{lutheremery} we show that along the so-called
Luther-Emery line, where the interacting electronic problem is 
exactly solvable, that the electron-phonon
coupling reduces to the problem of polaron dressing of the soliton
excitations (the polaronic soliton); Section \ref{conclusions}
contains our conclusions and a discussion of the interplay between
electron-electron interaction and the lattice potential.

\section{Description of the Model}
\label{model}

Our system of interest can be broken into three different pieces:
electrons, phonons and pseudo-spins (two-level systems). The electrons
are described in term of creation, $\psi^{\dag}_{\sigma}(x)$,
and annihilation, $\psi_{\sigma}(x)$, operators at position $x$ with
spin $\sigma$ ($\sigma = \uparrow,\downarrow$) and obey anti-commutation
relations:
$
\left\{ \psi_{\sigma} \left(x\right) ,\psi_{\sigma'}^{\dagger }
\left( x^{\prime }\right)
\right\} =\delta \left( x-x^{\prime }\right) \delta_{\sigma,\sigma'} 
$.
In the absence of interactions the Hamiltonian describing the
electron motion is simply:
\begin{equation}
H_{0}=\int dx \sum_{\sigma} \left\{ \frac{\hbar^{2}}{2m}
\frac{\partial \psi_{\sigma}^{\dagger }}{\partial
x}\frac{\partial \psi_{\sigma} }{\partial x} + V(x) \psi^{\dagger}_{\sigma}(x)
\psi_{\sigma}(x) \right\} \, ,
\label{band hamiltonian}
\end{equation}
where $m$ is the electron mass and $V(x)$ is the lattice potential.
If $V(x)$ is periodic then a gap opens up in the spectrum and if the
number of electrons per unit cell is $2$ the problem is described
as a band insulator. 

While the above description is satisfactory for a system where the electronic
wave-function is extended (and therefore the system is well-described
by an electron gas), in systems where the coupling with the ions is strong
a tight-binding description is usually a better starting point. 
In fact this kind of description is the starting point for the description
of the Mott insulator.
The electron Hamiltonian including the electron-electron
interactions can be generically written as:
\begin{equation}
H_{e}=- t_e \sum\limits_{i,\sigma }C_{i,\sigma }^{+}C_{i+1,\sigma
}+U\sum\limits_{i}n_{i,\uparrow }n_{i,\downarrow }+V\sum\limits_{i,\sigma
,\sigma ^{\prime }}n_{i,\sigma }n_{i+1,\sigma ^{\prime }} \, ,
\label{Extended Hubbard Model}
\end{equation}
where $C_{i,\sigma }$ ($C_{i,\sigma }^{+}$) is the annihilation (creation)
electron operator at unit cell $i$ with spin $\sigma $ and $n_{i,\sigma
}=C_{i,\sigma }^{+}C_{i,\sigma }$ is the electron number operator. 
$t_e$ is the hopping energy of the electron between different sites
and $U$ and $V
$ describe the on-site and the nearest neighbor electron-electron
interactions, respectively. 

In the non-interacting limit the electron energy is $E_k = - 2 t_e
\cos(k a)$, where $a$ is the lattice spacing. For a finite density of
electrons all the states up to the Fermi energy, $E_F$, are filled
generating two Fermi points defined by 
$E_{\pm k_F} = E_F$, where $k_F = \pi n/2$ is the Fermi momentum
($n$ is the number of electrons per unit of length). 
Close to the Fermi points the spectrum is $E_{k_F+q} -E_F = \upsilon_F q$ 
where $\upsilon_{F}=2 t_e a \sin \left(k_{F}a\right)$, is the Fermi velocity.
The linearization of the spectrum close to the Fermi points can also 
be translated into operator language. We rewrite the electron
operator as the product of a rapidly varying part ($e^{i \pm k_F x}$) and a 
slowly varying part:
\begin{eqnarray}
\psi_{\sigma}(x) = \psi_{R,\sigma}(x) e^{i k_F x}
+ \psi_{L,\sigma}(x) e^{-i k_F x} \, \, ,
\label{slowf}
\end{eqnarray}
where $\psi_{R,\sigma}(x)$ ($\psi_{L,\sigma}(x)$) creates a right (left)
moving electron in the system. Because of the chiral nature of these
excitations they can be described in terms of bosonic operators \cite{Voit}:
\begin{eqnarray}
\psi_{(R,L),\sigma}(s) = \frac{1}{\sqrt{2 \pi a}} e^{\pm i \sqrt{\pi}
\phi_{(R,L),\sigma}(s)} \, .
\label{boso1}
\end{eqnarray}
In turn, the bosonic modes $\phi_{(R,L),\sigma}$ 
can be described in terms of new fields, 
$\phi_{\sigma}$ and $\theta_{\sigma}$, as
$
\phi_{(R,L),\sigma}(x) = \phi_{\sigma}(x) \mp \theta_{\sigma}(x) .
$
The bosonic fields can then 
be rewritten in terms of charge and spin bosonic modes:
\begin{eqnarray}
\Phi_{\rho,\sigma} &=& \frac{1}{\sqrt{2}} \left(\phi_{\uparrow} \pm
\phi_{\downarrow}\right)
\nonumber
\\
\Theta_{\rho,\sigma} &=& \frac{1}{\sqrt{2}} \left(\theta_{\uparrow} \pm
\theta_{\downarrow}\right) \, \, .
\label{bosonization}
\end{eqnarray}
The procedure of writing electron operators in terms of bosons is called
bosonization.
It can be shown that the charge density operator 
$\rho \left( x\right) $ is related to 
$\Phi_{\rho }\left( x\right) $ by 
\begin{equation}
\rho \left( x\right) =-\frac{\sqrt{2}}{\pi }\frac{\partial \Phi _{\rho
}\left( x\right) }{\partial x}.  
\label{charge density operator}
\end{equation}
The non-interacting Hamiltonian and all the scattering process in the
1D system can now be rewritten in terms of bosons. There are three main
types of scattering in 1D: forward scattering (with small momentum transfer),
backscattering (with $2 k_F$ momentum transfer) and Umklapp scattering
(with $4 k_F$ momentum transfer). While the forward and backscattering do
not require the lattice to participate, the Umklapp scattering requires,
by momentum conservation, that
$4 k_F = G$ where $G = 2 \pi/a$ is the reciprocal lattice vector. Because
of this constraint the Umklapp scattering is only important at commensurate
filling factors and is responsible for the opening of the Mott gap
in the half-filled system ($n=1/a$).

The model (\ref{Extended Hubbard Model}) can be bosonized using the fields
$\Phi _{\rho }\left( x\right) $\ and $\Phi _{\sigma }\left(
x\right)$. The Hamiltonian of the problem breaks into
two main parts, $H=H_s+H_c$, 
describing spin $H_s$ and charge $H_c$ \cite{Voit}:
\begin{eqnarray}
H_s &=& H_{\sigma }+\frac{2g_{1}}{\left( 2\pi a\right) ^{2}}\int
dx\cos \left[ \sqrt{8}\Phi _{\sigma }\left( x\right) \right] 
\nonumber
\\
H_c &=& H_{\rho } + \frac{2g_{3}}{%
\left( 2\pi a\right) ^{2}}\int dx\cos \left[ \sqrt{8}\Phi _{\rho }\left(
x\right) +4k_{F}x\right] \, \, .
\label{bosonized hamiltonian} 
\end{eqnarray}
In the above Hamiltonian we have
\begin{eqnarray}
H_{v}=\frac{1}{2\pi }\int dx\left[ \left( u_{v}K_{v}\right) \pi ^{2}\Pi
_{v}^{2}\left( x\right) +\frac{u_{v}}{K_{v}}\left( \frac{\partial \Phi
_{v}\left( x\right) }{\partial x}\right)^{2}\right] ,\qquad v=\rho ,\sigma 
\label{luttinger}
\end{eqnarray}
describing a gapless Luttinger liquid involving only the forward
scattering processes. 
Here, $u_{v}$ are the velocity of the bosonic fields, $K_{v}$ are
the Luttinger parameters, and 
$\Pi _{v}\left( x\right) $ is the momentum field operator that is
canonically conjugate to $\Phi _{v}\left( x\right) $ : $\left[ \Phi
_{v}\left( x\right) ,\Pi _{v^{\prime }}\left( x^{\prime }\right) \right]
=i\delta _{v,v^{\prime }}\delta \left( x-x^{\prime }\right) $. 

In (\ref{bosonized hamiltonian}) the coupling constants $g_{1}$
and $g_{3}$ represent the backward scattering and the
Umklapp process. 
One can relate the parameters in the bosonized Hamiltonian with
the ones in (\ref{Extended Hubbard Model}) by \cite{Emery}:
\begin{eqnarray}
u_{\rho }K_{\rho } &=&u_{\sigma }K_{\sigma }=\upsilon _{F}-\frac{Va}{\pi } 
\nonumber \\
u_{\rho }/K_{\rho } &=&\upsilon _{F}+\frac{\left( U+5V\right) a}{\pi } 
\nonumber \\
u_{\sigma }/K_{\sigma } &=&\upsilon _{F}-\frac{\left( U-V\right) a}{\pi } 
\nonumber \\
\frac{g_{1}}{a} &=&\frac{g_{3}}{a}=U-2V.  
\label{u & K & g1 & g3} \, .
\end{eqnarray}
This relation between the coupling constants in the bosonic theory and
the interactions in the electron problem allow us to change freely from one
representation to the other.
Notice that the bosonization procedure is not a good starting point for
the problem of the band insulator because $n=2/a$ and therefore 
$\upsilon_F = 0$, implying that we cannot define right and left moving
particles (this is just the trivial statement that when the band is full
there is no electronic motion). 

Observe that the bosonic fields in (\ref{bosonized hamiltonian}) separate
completely into the spin and charge parts. This is called spin-charge
separation and it is a generic property of one dimensional systems. Namely,
while an isolated electron carries both spin and charge, in the 1D system
the electron decays into the bosonic collective modes described above that
carry spin and charge separately. Since we are only considering the properties
of insulators, the spin degrees of freedom decouple from the problem and
are completely described by $H_s$ in (\ref{bosonized hamiltonian}). 
From this result we can immediately draw some conclusions about the nature
of the spin excitations in these systems. 

The simplest way to understand
the importance of these terms is to perform a perturbative 
renormalization group (RG) calculation by assuming that the non-linear
terms in (\ref{bosonized hamiltonian}) are small compared to the
Luttinger liquid terms in (\ref{luttinger}). We can
study how these non-linear terms affect the physics of the Luttinger
liquid. The RG procedure is defined by introducing a cut-off $\Lambda$
in momentum space so that all the momenta, $k$, are defined in 
$-\Lambda < k < \Lambda$. We now trace all degrees of freedom
in the region of high momenta around the cut-off, say, $\Lambda/b<|k|<\Lambda$
($b>1$),
and rescale the frequencies and momenta so that the Luttinger liquid
Hamiltonian (\ref{luttinger}) becomes invariant under rescaling
($\omega \to b \omega$
and $k \to b k$). As we rescale the system to lower energies the
coupling constants change with the scale. This change is given by
the RG equations. For the backscattering problem it can be shown
that \cite{Voit}:
\begin{eqnarray}
\frac{\partial g_1(\ell)}{\partial \ell} &=& 2 (1-K_{\sigma}) g_1(\ell) \, ,
\label{rg1}
\end{eqnarray}
where $\ell = \ln(1/\Lambda)$.  
Notice that for $K_{\sigma}>1$, the backscattering
term scales to zero and therefore it is irrelevant under the RG.
The spin degrees of freedom are completely described by 
the Luttinger liquid Hamiltonian (\ref{luttinger}). The
spin excitations are gapless and propagate with velocity $v_{\sigma}$.
However, when $K_{\sigma}<1$, the backscattering
term grows under the RG, implying that the non-linear terms are becoming
stronger at low energies: the backscattering
is relevant. The relevance of this term can be easily understood if
we consider only the non-linear part of the Hamiltonian:
\begin{eqnarray}
H_I &=&  \frac{g_1}{2 (\pi a)^2} \int dx \, \, 
\cos(\sqrt{8} (\Phi_{\sigma}(x)) \, .
\end{eqnarray}
If $g_1 \to \infty$ at long length scales we see that the coefficient
of the cosine grows and becomes much bigger than the energy scale of the 
Luttinger liquid in (\ref{luttinger}). 
If we now ignore completely the Luttinger liquid
part of the Hamiltonian (as the RG procedure tells us we should do), 
we see that the energy of the problem is minimized when the field
$\Phi_{\sigma}(x) = 0$ for $g_1<0$ or $\Phi_{\sigma}(x) = 
\sqrt{\pi/8}$ when $g_1>0$.
Thus, in the ground state the field is uniform. Let us
allow the field $\Phi_{\sigma}$ to fluctuate around its mimimum value. 
Expanding $H_I$ up to second order in $\Phi_{\sigma}$ we find
\begin{eqnarray}
H_I \approx - \frac{|g_1| L}{2 (\pi a)^2} + \frac{2 |g_1|}{(\pi a)^2}
\int dx \Phi_{\sigma}^2(x) \, \, ,
\label{hi}
\end{eqnarray}
where $L$ is the size of the system. The first term in 
(\ref{hi}) is the ground state energy of the uniform state. 
Adding now the Luttinger liquid part
of the Hamiltonian, Eq.(\ref{luttinger}), we find that the action
that describes the system is:
\begin{eqnarray}
S &\approx& -  \frac{|g_1| L}{2 (\pi a)^2} +  
\int dx \int d\tau
\left[\frac{1}{2 K_{\sigma} \upsilon_{\sigma}} 
\left(\partial_{\tau} \Phi_{\sigma}(x) \right)^2 + 
\frac{\upsilon_{\sigma}}{2 K_{\sigma}} \left(\partial_{x} 
\Phi_{\sigma}\right)^2
+  \frac{2 |g_1|}{(\pi a)^2} \Phi_{\sigma}^2(x)
\right]
\nonumber
\\
&\approx& -  \frac{|g_1| L}{2 (\pi a)^2} +  
\int \frac{dq}{2 \pi} \int \frac{d\omega}{2 \pi}
\left[\frac{1}{2 K_{\sigma} \upsilon_{\sigma}} \omega^2 + 
\frac{\upsilon_{\sigma}}{2 K_{\sigma}} 
q^2 + \frac{2 |g_1|}{(\pi a)^2}\right] |\Phi_{\sigma}(q,\omega)|^2 \, ,
\end{eqnarray}
where we have Fourier transformed the fields in the last line.
Notice that the excitation spectrum has changed from $\omega(q) = 
\pm \upsilon_{\sigma} |q|$ to 
\begin{eqnarray}
\omega(q) &=& \pm \upsilon_{\sigma} \sqrt{q^2 + m^2}
\nonumber
\\ 
m &=& 
\frac{2}{\pi a} \sqrt{K_{\sigma} |g_1| \upsilon_{\sigma}} \, .
\end{eqnarray}
This result indicates that a gap of energy 
\begin{eqnarray}
\Delta_S = \upsilon_{\sigma} m = 2 \sqrt{\frac{K_{\sigma} \upsilon_{\sigma} 
|g_1|}{\pi a}}
\label{gap}
\end{eqnarray} 
opens in the spectrum at $q=0$ ($k=k_F$ in terms of the original
electrons). We can define
a correlation lenght $\xi_{\sigma} = 1/m = v_{\sigma}/\Delta_0$
which tells us that the correlations decay exponentially over $\xi_{\sigma}$
and that for distances larger than $\xi_{\sigma}$ no correlations are possible.
Thus, the relevance of the operators
under the RG implies that the low lying spin excitations of the system are
gapped (in contrast to the Luttinger liquid).
This is the so-called spin gap and 
is a sign of Cooper pairing in the system since for $K_{\sigma}<1$ the
interactions become effectively attractive. To understand the origin
of this result we notice that
when two electrons form a Cooper pair in a singlet state it costs energy
to break the pair and make a spin excitation. 

Let us focus on the charge part of the Hamiltonian that can
be written as:
\begin{equation}
H_{e}=\frac{1}{2\pi }\int dx\left[ \left( u_{\rho }K_{\rho }\right) \pi
^{2}\Pi _{\rho }^{2}+\frac{u_{\rho }}{K_{\rho }}\left( \frac{\partial \Phi
_{\rho }}{\partial x}\right) ^{2}\right] +\frac{2g_{3}}{\left( 2\pi \alpha
\right) ^{2}}\int dx\cos \left[ \sqrt{8}\Phi _{\rho } + 4 k_F x\right] \, .
\label{mott insulator}
\end{equation}
The problem described by (\ref{mott insulator}) is well understood and
it can be shown that when $4 k_F a \neq 2 \pi$ the Umklapp term is
irrelevant and the problem is described by a gapless Luttinger liquid.
However, when the system becomes commensurate with the lattice ($n=1/a$)
the Umklapp term in (\ref{mott insulator}) is important. In fact, an RG
calculation gives \cite{Voit}:
\begin{eqnarray}
\frac{\partial g_3(\ell)}{\partial \ell} &=& 2 (1-K_{\rho}) g_3(\ell) \, .
\label{rg2}
\end{eqnarray}
In complete analogy with the backscattering problem, this result shows
that a gap opens in the charge spectrum (the so-called Mott gap) 
if $K_{\rho}<1$. The Umklapp term 
is irrelevant when $K_{\rho}>1$. Thus, a correlated insulator has two major 
requirements. Namely, the electron-electron interactions have to be repulsive
and the charge density has to be commensurate with the lattice.

In fact the calculation of the gap in (\ref{gap}) is incorrect because
we have not taken into account the fluctuations in the problem by
expanding the fields close to its minimum. We can learn more about 
the {\it actual} value of the Mott gap by looking at the RG equation
(\ref{rg2}) more closely. That RG predicted that for 
$K_{\rho}<1$ the system scales to strong coupling
and $g_3$ grows without bound. The maximum value $g_3$ can attain in our
theory is the largest energy scale, that is, the electron bandwidth, 
$W \approx 4 t_e$. 
Beyond this energy scale the perturbative RG calculation breaksdown.
In our linearized theory it is easy to see that $W \approx v_F \Lambda_0$,
where $\Lambda_0 \approx 1/a$ is the bare cut-off in the theory.
When $g_3(\ell)$ becomes of the order of $W$ the variable $\ell$
reaches its maximum value $\ell^* = \ln(\Lambda_0/\Lambda^*)$.
Here $\Lambda^*$, which has dimensions of inverse length, becomes
of the order of the inverse of correlation length, that is, 
$\Lambda^* \approx 1/\xi_{\rho} \approx \Delta/v_{\rho}$. 
However, accordingly to (\ref{rg2})
we must have
\begin{eqnarray}
\ell^* &=& \frac{1}{2 (1-K_{\rho})} \ln(g(\ell^*)/g_3(0))
\nonumber
\\
\ln(W/\Delta) &\approx& \frac{1}{2 (1-K_{\rho})} \ln(W/|g_3|) \, ,
\end{eqnarray}
which can be solved for $\Delta$ as
\begin{eqnarray}
\Delta_M = W \left(\frac{|g_3|}{W}\right)^{\frac{1}{2(1-K_{\rho})}}
\label{delta}
\end{eqnarray}
showing that the gap is proportional to $|g_3|^{1/(2(1-K_{\rho}))}$ 
instead of $|g_3|^{1/2}$ as predicted by our naive expansion (\ref{gap}).
In fact the square root behavior of the
gap with the coupling constant 
is only obtained when $K_{\rho}=0$ which is equivalent
to the strong coupling regime where the interaction becomes
of the order of the bandwidth of the original problem 
(see (\ref{u & K & g1 & g3})). The exact value of $\Delta_M$ 
can be obtained by the mapping, via a Jordan-Wigner
transformation, the Umklapp problem into the anisotropic Heisenberg model 
\cite{fradkin}. This model can be solved exactly by Bethe ansatz and the
gap calculated without approximations\cite{Luther}. It is found that
the power law dependence of the Mott gap with the coupling constant
in (\ref{delta}) is exact. This result shows the importance of the RG
calculation. Observe that the bandwidth $W$ which appears in 
(\ref{delta}) is a non-universal number that cannot be obtained from
the RG. By comparing (\ref{delta}) with (\ref{gap}), however, we find
$W \approx 4 v_F/(\pi a)$.

The other important components of the problem are the phonons. These can
be of two types: acoustic or optical. Since we are only interested in
the low energy physics of the 1D system we consider only the acoustic
modes that can be described by the Hamiltonian \cite{Kittel} 
\begin{equation}
H_{ph}=\frac{1}{2}\int dx\left[ \frac{\Pi_L^{2}\left( x\right) }{\rho_L%
}+c_{s}^{2}\rho_L\left( \frac{\partial \phi_L\left( x\right) }{%
\partial x}\right) ^{2}\right] ,  \label{phonon hamiltonian}
\end{equation}
where $\rho_L$ is the mass density of the system, $c_s$ is the sound
velocity, and  
$\Pi_L\left( x\right)$ is the phonon momentum operator that is 
canonically conjugated to the phonon field $\phi_L\left( x\right)$:
$\left[ \phi_L\left( x\right) ,\Pi_L\left( x^{\prime }\right) \right]
=i\hbar \delta \left( x-x^{\prime }\right)$. 

The electron-phonon coupling assumed here is of the deformation potential
type, which is appropriate for acoustic phonons:
\begin{eqnarray}
H_{e-p} &=&\gamma _{p}\int dx \, \, \rho \left( x\right) 
\frac{\partial \phi_{L}\left( x\right) }{\partial x}  \nonumber \\
&=&\gamma _{p}\int dx \, \, \psi ^{\dagger }\left( x\right) 
\psi \left( x\right) 
\frac{\partial \phi_L\left( x\right) }{\partial x}
\nonumber
\\
&=& \frac{\gamma_p \sqrt{2}}{\pi} \int dx
\, \frac{\partial \Phi _{\rho }\left( x\right) }{\partial
x}\frac{\partial \phi_L\left( x\right) }{\partial x} \, .
\label{band insulator electron-phonon interaction}
\end{eqnarray}

Another component in our problem is the presence of pseudo-spins (or two-level
systems) that couple to the electronic 
charge degrees of freedom but not to the spins (this is why we term them
pseudo-spins). In a real system these pseudo-spins can be dangling bonds of
atoms in a nano-structure or hydrogen bonds in DNA. Because of
their nature the pseudospins are assumed to be non-interacting and 
are described by a Hamiltonian:
\begin{eqnarray}
H_{t} = \frac{1}{2}
\int dx\left[ -\varepsilon \sigma ^{z}(x)+ t \sigma ^{x}(x)
\right] ,  
\label{two-level systems hamiltonian}
\end{eqnarray}
where $\sigma ^{z}$ and $\sigma ^{x}$ are Pauli matrices
that obey: $\left[ \sigma ^{z}(x),\sigma ^{x}(x^{\prime })\right] =2i\sigma
^{y}(x)\delta \left( x-x^{\prime }\right)$.
In (\ref{two-level systems hamiltonian}) 
$\varepsilon > 0$ is the energy difference between the ground
state with pseudo-spin $\uparrow$ and the first excited state
with pseudo-spin $\downarrow$. For generality we also allow
for the quantum tunneling of the pseudo-spin between the two
possible configurations and $t$ is the energy associated
with this tunneling. 

As mentioned previously, the coupling between the pseudo-spins
and the charge degrees of freedom has essentially the same structure
as the electron-phonon coupling:
\begin{eqnarray}
H_{e-t} &=& \frac{\gamma _{t}}{2}\int dx\psi ^{\dagger }\left(
x\right) \psi \left( x\right) \left[ \sigma ^{z}(x)-1\right]
\nonumber
\\
&=& \frac{\gamma_t}{\sqrt{2} \pi }\int dx\frac{\partial \Phi _{\rho
}\left( x\right) }{\partial x}\left[ \sigma ^{z}(x)-1\right] \, .
\label{het}
\end{eqnarray}
The rationale for this coupling is the fact that electron
polarization around a pseudo-spin can lead to its local reorientation. 
This is the same reasoning behind the electron-phonon
coupling except that in the case of the pseudo-spin the reorientation
process is quantized.

In order to simplify the problem we project out the pseudo-spin
variables in terms of a new field $\theta(x)$ by writing the
generic form for the pseudo-spin wavefunction, 
\begin{eqnarray}
\Psi(x) = \left( \begin{array}{cc} \cos \theta(x) \\ \sin \theta(x) 
\end{array} \right) \, ,
\end{eqnarray}
and rewriting the Hamiltonian as:
\begin{eqnarray}
H_t \to \langle \Psi|H_{t}|\Psi \rangle 
&=&\frac{1}{2}\int dx\left[ -\varepsilon \cos 2\theta (x)+t\sin 2\theta (x)
\right]  
\nonumber
\\
H_{e-t} \to \langle \Psi|H_{e-t}|\Psi \rangle 
&=&-\gamma_{t}\int dx \psi ^{\dagger }(x) \psi(x) \sin ^{2}\theta \left(
x\right) .
\end{eqnarray}
Now, all the variables in the problem are given in terms of fields that
are continuous and smoothly varying in space.

\section{Polarons in Band Insulators}
\label{band}

In this section we review the problem of polarons in band insulators
described by the Hamiltonian (\ref{band hamiltonian}). 
The problem reduces to a valence band which is full and a conduction
band that is empty.
In this case, as discussed
previously, the bosonization procedure is not a good starting point since
the charge density is such that the Fermi velocity vanishes and therefore
the linearization around the Fermi point is a bad approximation.
The problem of bosonization
and band insulators for weak periodic potentials is discussed in Section
\ref{conclusions}. In this section we disregard electron-electron 
interactions and consider the problem of a single electron (hole) at
the bottom (top) of the conduction (valence) band. 
The Hamiltonian for the electron in the conduction band
is identical to the one in (\ref{band hamiltonian}) 
except that we can set $V=0$
and replace the electron bare mass $m$ by its effective band mass $m_c$.
The full Hamiltonian of the problem is written as
$H_{p} = H_{0}+H_{ph}+H_{t}+H_{e-p}+H_{e-t}$ where:
\begin{eqnarray}
H_{0} &=&\frac{\hbar ^{2}}{2 m_c}\int dx
\frac{\partial \psi ^{\dagger }}{\partial x}
\frac{\partial \psi }{\partial x}  
\nonumber \\
H_{ph} &=& \frac{1}{2}\int dx\left[ \frac{\Pi_L^{2}}{\rho_L}
+c_{s}^{2}\rho_L \left( \frac{\partial \phi_L}{\partial x}\right)^{2}
\right]   
\nonumber \\
H_{t} &=& 
\frac{1}{2}\int dx\left[ -\varepsilon \cos 2\theta (x)+t\sin 2\theta (x)
\right] 
\nonumber \\
H_{e-p} &=& \gamma _{p}\int dx\psi ^{\dagger }\psi \frac{\partial \phi_L
}{\partial x}  
\nonumber \\
H_{e-t} &=&
-\gamma _{t}\int dx \psi ^{\dagger }(x) \psi(x) \sin ^{2}\theta \left(
x\right) .
\label{band insulator total Hamiltonian}
\end{eqnarray}

The generating function of the problem can be written as:
\begin{eqnarray}
Z = \int D\psi^* D \psi D\phi_L D\theta \, \, e^{- \frac{i}{\hbar} 
\int dt \int dx {\cal L}[\psi^*,\psi,\phi_L,\theta]} \, \, ,
\label{z}
\end{eqnarray}
where $\psi^*$ and $\psi$ are Grassmann variables and the Lagrangean
of the problem is:
\begin{eqnarray}
{\cal L} &=& i\hbar \left( \psi ^{\ast }\frac{\partial \psi }{\partial t}-\psi 
\frac{\partial \psi ^{\ast }}{\partial t}\right) 
-\frac{\hbar ^{2}}{2 m_c} \left( \frac{\partial \psi ^{\ast }}{\partial x}\right) \left( \frac{ \partial \psi }{\partial x}\right) 
+\frac{\rho_L}{2}\left[ \left( \frac{\partial \phi_L}{\partial t}\right)^{2}
-c_{s}^{2}\left( \frac{\partial \phi_L}{\partial x}\right) ^{2}\right]   
\nonumber \\
&+& \frac{1}{2}\left[ \varepsilon \cos 2\theta -t\sin 2\theta \right] 
-\left[\gamma_{p} \frac{\partial \phi_L}{\partial x}
-\gamma_{t}\sin ^{2}\theta(x)\right]|\psi(x)|^2 .
\label{band insulator Lagrangean}
\end{eqnarray}
Although the above Lagrangean is of relative complexity we are going
to study the problem in the semiclassical limit by letting $\hbar \to 0$. 
In this limit, as one can see directly from (\ref{z}), the problem
is dominated by the saddle point equations:
\begin{eqnarray}
t\cos 2\theta  &=&-\varepsilon \sin 2\theta +\gamma _{b\sigma }\left| \psi
\right| ^{2}\sin 2\theta   
\nonumber \\
i\hbar \frac{\partial \psi }{\partial t} &=&-\frac{\hbar ^{2}}{2m_c}
\frac{\partial ^{2}\psi}{\partial ^{2}x}+
\left( \gamma _{p}\frac{\partial \phi_{L}}{\partial x}-
\gamma _{t}\sin ^{2}\theta \right) \psi(x)   \nonumber
\\
\rho_L\frac{\partial ^{2}\phi_L}{\partial t^{2}} &=&c_{s}^{2}\rho_L%
\frac{\partial ^{2}\phi_L}{\partial x^{2}}+\gamma _{p}\frac{\partial
\left| \psi \right| ^{2}}{\partial x}.
\label{band insulator equations}
\end{eqnarray}
The first equation in (\ref{band insulator equations}) gives: 
\begin{eqnarray}
\tan 2\theta(x) =\frac{t}{\gamma _{t}\left| \psi(x) \right|^{2}-\varepsilon } \, ,
\label{equation for theta}
\end{eqnarray}
which shows that the pseudo-spins can rearrange their orientation
due to the presence of the electrons. In fact if we define a position
dependent pseudo-spin energy:
\begin{eqnarray}
\bar{\varepsilon}(x) = \varepsilon-\gamma_t |\psi(x)|^2
\label{renormalized stiffness}
\end{eqnarray}
we see that the electrons decrease the effective pseudo-spin energy
allowing for the tunneling of the pseudo-spins. In other words, the
electrons delocalize the pseudo-spins. In what follows we are going
to consider two regimes. In the first regime 
we have $t \gg \bar{\varepsilon}(x)$
and therefore the pseudospins delocalize, in the second regime 
$t \ll \bar{\varepsilon}$
the renormalization of the pseudo-spin energy is not enough to delocalize
the pseudo-spins. As we are going to see, in both limits the physics is
very similar and therefore the final behavior of the polaron is 
essentially independent of these limits. The expression (\ref{equation for
theta}) can also be written as:
\begin{eqnarray}
\sin^{2}\theta(x) =  \frac{1}{2}-\frac{\bar{\varepsilon}(x)}{
2\sqrt{\bar{\varepsilon}^{2}(x)+ t^{2}}} \, .
\end{eqnarray}
For $t \gg \bar{\varepsilon}$ we have
\begin{eqnarray}
\sin^{2}\theta(x) \approx
\frac{1}{2t}\left(\gamma _{t}\left| \psi(x) \right|^{2}-\varepsilon
+t\right) \, \, ,
\label{limits0}
\end{eqnarray}
while for $t \ll \bar{\varepsilon}$ we have
\begin{eqnarray}
\sin^{2}\theta(x) \approx 
\frac{t^{2}}{2\varepsilon^{3}} \left(\gamma_{t} \left|\psi(x)\right|^{2}
+\frac{\varepsilon}{2}\right) \, .
\label{limits}
\end{eqnarray}

Substituting (\ref{limits0}) or 
(\ref{limits}) back into (\ref{band insulator equations})
we find:
\begin{eqnarray}
i\hbar \frac{\partial \psi }{\partial t} &=&-\frac{\hbar ^{2}}{2m}\frac{%
\partial ^{2}\psi }{\partial ^{2}x}+\gamma _{p}\frac{\partial \phi_L}{%
\partial x}\psi -\left( D_{t}\left| \psi \right| ^{2}+E_{t}\right) \psi  
\nonumber
\\
\rho_L\frac{\partial ^{2}\phi_L}{\partial t^{2}} &=&c_{s}^{2}\rho_L%
\frac{\partial ^{2}\phi_L}{\partial x^{2}}+\gamma _{p}\frac{\partial
\left| \psi \right| ^{2}}{\partial x} \, ,
\label{davydov}
\end{eqnarray}
where the values of $D_{t}$ and $E_{t}$ are giving by
\begin{eqnarray}
D_{t} &\approx& \frac{1}{2t}\gamma _{t}^{2}
\nonumber
\\
E_{t} &\approx& \frac{1}{2}\left( 1-\frac{\varepsilon }{t}\right) \gamma _{t} 
\end{eqnarray}
for $t \gg \bar{\varepsilon}$, and
\begin{eqnarray}
D_{t} &\approx&  \frac{t^{2}}{2\varepsilon^{3}}\gamma^2_{t}
\nonumber \\
E_{t} &\approx& \frac{t^{2}}{4\varepsilon^{2}}\gamma _{t}
\label{A and B}
\end{eqnarray}
for $t \ll \bar{\varepsilon}$. The form of the equations in (\ref{davydov})
is similar to the ones studied by Davydov \cite{dav} in the
problem of propagations of solitonic waves in biological systems by
taken into account electrons and phonons. This should not be surprising
since they describe non-interacting electrons interacting with harmonic
excitations. We should stress, however, that our equations also include
the pseudo-spin variables that were not considered in ref.\cite{dav}.

The solution of the problem can now be obtained by noticing that
both the electron field, $\psi$, and the phonon field, $\phi_L$, 
have the form of a traveling wave. Using the Galilean invariance
of the problem we write:
\begin{eqnarray}
\psi \left( x,t\right) &=& 
\phi _{0}\left(x \mp \upsilon t \right) \exp \left[ \frac{i}{%
\hbar }\left( m\upsilon x-E_{0}t\right) \right] ,
\nonumber
\\
\phi_{L}\left(x,t\right) &=& \phi_L\left(x \mp \upsilon t\right) =
\phi_L\left(\lambda \right) \, \, ,
\end{eqnarray}
where $\lambda =x \mp \upsilon t$. Here $E_0$ is the binding energy of the
electron and $\upsilon $ is the
speed of the traveling wave. After this simple transformation we find:
\begin{eqnarray}
-\frac{\hbar^{2}}{2m_c}\frac{d^{2}\phi _{0}}{d\lambda ^{2}}+
\gamma _{p}\frac{d\phi_L}{d\lambda }\phi _{0} &=&
\left[ E_{0}-\frac{m_c \upsilon^{2}}{2}
+E_{t}\right] \phi _{0}+D_{t}\phi _{0}^{3}  
\nonumber 
\\
\rho_L\left( \upsilon ^{2}-c_{s}^{2}\right) \frac{d\phi_L}{d\lambda }
&=&\gamma _{p}\phi _{0}^{2}  \, .
\label{band insulator soliton equations}
\end{eqnarray}
Combining both equations, we have 
\begin{equation}
-\frac{\hbar^{2}}{2m_c}\frac{d^{2}\phi _{0}}{d\lambda ^{2}}=\left[E_{0}-
\frac{m\upsilon ^{2}}{2}+E_{t}\right] \phi _{0}+\left[ D_{t}+
\frac{\gamma_{p}^{2}}{\rho_L\left( c_{s}^{2}-\upsilon ^{2}\right) }%
\right] \phi _{0}^{3} \, ,
\end{equation}
which is the non-linear Schr\"{o}dinger equation supporting the
propagation of a soliton in the system. This soliton is nothing
but a polaron formed of the polarization of the lattice and
the pseudo-spins. It is simple to show that for one electron (or hole)
the solution is:
\begin{equation}
\phi_{0}(x,t) =\frac{\sqrt{g}}{2}\text{sech}\left[ \frac{g\left( x-\upsilon
t\right) }{2}\right] ,  
\label{band insulator solution for electron}
\end{equation}
where 
\begin{equation}
g=\frac{m_c}{\hbar ^{2}}\left[D_{t}+\frac{\gamma _{p}^{2}}{\rho_{L}
\left( c_{s}^{2}-\upsilon ^{2}\right) }\right] 
\end{equation}
and the binding energy is
\begin{equation}
E_{0}=\frac{m_c\upsilon ^{2}}{2}-\frac{\hbar ^{2}g^{2}}{8m_c}-E_{t}.
\end{equation}
By comparing this result with the one found in reference 
\cite{Castro Neto} where
pseudo-spins are not considered, we notice the
binding energy is lowered by $E_{t}$. Thus, 
the soliton becomes more stable due to the electron coupling with
two-level systems.
Moreover, since the size of the polaron is of order $\xi \approx 1/g$
we see that the coupling to the pseudo-spins also decreases the size
of the polaron by a factor of $D_t$. The relative lattice displacement
is given by (\ref{band insulator soliton equations}) 
\begin{equation}
\phi_L(x,t)=-\frac{\gamma _{p}}{2\rho_L\left( c_{s}^{2}-\upsilon^{2}\right)}
\tanh \left[ \frac{g\left( x-\upsilon t\right) }{2}\right] .
\label{band insulator solution for phonon}
\end{equation}

\section{Polaronic Soliton in Mott insulators}
\label{mott}

In the previous section we studied the problem of polaron formation in
a band insulator. In this section we will study the problem of
non-linear excitations in a Mott insulator where the gap is generated by
correlations, not by the interaction with the lattice. Since this is
an interacting problem it is clearly more involved. However, because we
are considering a system where the electronic density is commensurate
with the lattice and the band is not full, we can use the tools of bosonization
to study it. The methods will be, however, very similar to the ones
used in the previous section. This will show that the polaronic soliton
formation in the Mott insulator is very similar to the problem of polaron
formation in a band insulator. 

The basic Hamiltonian of the problem is given by: 
$H_{m} = H_{ee}+H_{ph}+H_{t}+H_{e-p}+H_{e-t}$
where:
\begin{eqnarray}
H_{ee} &=& \frac{1}{2\pi }\int dx\left[ \left( u_{\rho }K_{\rho }\right) \pi
^{2}\Pi _{\rho }^{2}+\frac{u_{\rho }}{K_{\rho }}\left( \frac{\partial \Phi
_{\rho }}{\partial x}\right) ^{2}\right] +\frac{2g_{3}}{\left( 2\pi \alpha
\right) ^{2}}\int dx\cos \left[ \sqrt{8}\Phi _{\rho }\right]   
\nonumber 
\\
H_{ph} &=& \frac{1}{2}\int dx\left[ \frac{\Pi_L^{2}}{\rho_L}%
+c_{s}^{2}\rho_L\left( \frac{\partial \phi_L}{\partial x}\right) ^{2}%
\right]   
\nonumber 
\\
H_{t} &=& \frac{1}{2}\int dx\left[ -\varepsilon \cos 2\theta (x)+t\sin
2\theta (x)\right]   
\nonumber 
\\
H_{e-p} &=&\gamma _{p}\int dx\frac{\partial \Phi _{\rho }\left( x\right) 
}{\partial x}\frac{\partial \phi_L\left( x\right) }{\partial x}  
\nonumber
\\
H_{e-t} &=&- \gamma _{t}\int dx\frac{\partial \Phi _{\rho
}\left( x\right) }{\partial x}\sin ^{2}\theta \left( x\right) \, .
\label{mott insulator total Hamiltonian}
\end{eqnarray}

Once again the problem can be discussed in terms of the generating 
functional for the fields:
\begin{eqnarray}
Z = \int D\Phi_{\rho} D\phi_L D\theta \, \, 
e^{\frac{i}{\hbar} \int dt \int dx {\cal L}[\Phi_{\rho},\phi_L,\theta]} \, \, ,
\end{eqnarray}
where 
the Lagrangean of the system can then be written as 
\begin{eqnarray}
{\cal L} &=&
\frac{1}{2\pi }\left[ \frac{1}{u_{\rho }K_{\rho }}\left( \frac{
\partial \Phi _{\rho }}{\partial t}\right) ^{2}-\frac{u_{\rho }}{K_{\rho }}
\left( \frac{\partial \Phi _{\rho }}{\partial x}\right) ^{2}\right] 
-\frac{2g_{3}}{\left( 2\pi \alpha \right)^{2}}
\cos \left[ \sqrt{8}\Phi _{\rho }\right] 
+\frac{\rho_L}{2}\left[ \left( \frac{\partial \phi_L}{\partial t}\right)^{2}
-c_{s}^{2}\left( \frac{\partial \phi_L}{\partial x}\right)^{2}\right]   
\nonumber 
\\
&+& \frac{1}{2}\left[ \varepsilon \cos 2\theta -t\sin 2\theta \right] 
-\gamma_{p}\frac{\partial \Phi _{\rho }}{\partial x}
\frac{\partial \phi_L}{\partial x}
+\gamma _{t}\frac{\partial \Phi _{\rho }}{\partial x}\sin^{2}\theta   \, .
\label{mott insulator Lagrangean}
\end{eqnarray}
The equations of motion are:
\begin{eqnarray}
t\cos 2\theta  &=&-\varepsilon \sin 2\theta +
\gamma _{t}\frac{\partial \Phi _{\rho }}{\partial x}\sin 2\theta   
\nonumber 
\\
\frac{1}{u_{\rho }\pi K_{\rho }}\frac{\partial^{2}\Phi _{\rho }}{\partial
t^{2}} &=& \frac{u_{\rho }}{\pi K_{\rho }}
\frac{\partial ^{2}\Phi _{\rho }}{\partial x^{2}}
+\frac{\sqrt{2}g_{3}}{\left( \pi \alpha \right)^{2}}
\sin\left( \sqrt{8}\Phi _{\rho }\right) 
+\gamma _{p} \frac{\partial ^{2}\phi_L}{\partial x^{2}}
-\gamma _{t}\frac{\partial \sin ^{2}\theta }{\partial x}  
\nonumber 
\\
\rho_L\frac{\partial ^{2}\phi_L}{\partial t^{2}} &=& c_{s}^{2}\rho_L 
\frac{\partial^{2}\phi_L}{\partial x^{2}}+\gamma _{t}
\frac{\partial^{2}\Phi _{\rho }}{\partial x^{2}}.
\label{mott insulator Equations of motion}
\end{eqnarray}

The first equation in (\ref{mott insulator Equations of motion}) can
be solved exactly as in (\ref{equation for theta}) and
the effective local pseudo-spin energy is replaced by
\begin{eqnarray}
\bar{\varepsilon}(x)=\varepsilon -\gamma_{t} 
\frac{\partial \Phi _{\rho }}{\partial x} \, .
\end{eqnarray}
We consider the same limits as before, namely $t \gg \bar{\varepsilon}$
and $t \ll \bar{\varepsilon}$. The equations of motion for the fields are:
\begin{eqnarray}
\frac{1}{u_{\rho }\pi K_{\rho }}
\frac{\partial^{2}\Phi _{\rho }}{\partial t^{2}} &=&
\frac{u_{\rho}}{\pi K_{\rho}}\frac{\partial^{2}\Phi_{\rho }}{\partial x^{2}}
+\frac{\sqrt{2}g_{3}}{\left(\pi \alpha \right)^{2}}
\sin\left(\sqrt{8}\Phi _{\rho }\right) 
+\gamma _{p} \frac{\partial ^{2}\phi_L}{\partial x^{2}}
-D_{t}\frac{\partial ^{2}\Phi _{\rho }}{\partial x^{2}} 
\nonumber
\\
\rho_L \frac{\partial^{2}\phi_L}{\partial t^{2}} &=&
c_{s}^{2}\rho_L \frac{\partial^{2}\phi_L}{\partial x^{2}}
+\gamma_{p}\frac{\partial^{2}\Phi _{\rho }}{\partial x^{2}}.
\end{eqnarray}
This set of equations is very different from the ones obtained in
the case of the band insulator that is described by the Davydov's
equations (\ref{davydov}). In fact these equations contain non-perturbative
effects that cannot be described by the the non-interacting electron
picture. 

Using once again the Galilean invariance of the solutions we can write
$\Phi_{\rho }\left(x,t\right) =\Phi _{\rho }\left(x \mp \upsilon t\right) 
=\Phi_{\rho }\left(\lambda \right)$ and 
$\phi_{L}\left(x,t\right) =\phi_L\left(x \mp \upsilon t\right) 
=\phi_L\left(\lambda \right)$, where $\lambda =x \mp \upsilon t$.
The equations of motion reduce to:
\begin{eqnarray}
\left(\frac{\upsilon^{2}}{u_{\rho }\pi K_{\rho }}-
\frac{u_{\rho}}{\pi K_{\rho }}
+D_{t}\right) \frac{d^{2}\Phi_{\rho}}{d\lambda^{2}} &=&
\frac{\sqrt{2}g_{3}}{\left( \pi \alpha \right)^{2}}
\sin \left(\sqrt{8}\Phi_{\rho }\right) +\gamma_{p}
\frac{d^{2}\phi_L}{d\lambda^{2}}  
\nonumber
\\
\rho_L\left(\upsilon^{2}-c_{s}^{2}\right) \frac{d^{2}\phi_L}{d\lambda ^{2}} 
&=&\gamma_{p}\frac{d^{2}\Phi _{\rho}}{d\lambda ^{2}}.
\label{mott insulator soliton equations}
\end{eqnarray}
Combining both equations and replacing $\Phi _{\rho}$\ with $\Phi =\sqrt{8}%
\Phi_{\rho}$, the equation of motion for the charge field becomes: 
\begin{eqnarray}
\frac{d^{2}\Phi }{d\bar{\lambda}^{2}}= sign(\mu) \sin(\Phi(\bar{\lambda})) ,
\label{sinegordon}
\end{eqnarray}
where $sign(x) = +1 \, \, (-1)$ if $x>0 \, \, (x<0)$ is a sign function, 
\begin{equation}
\bar{\lambda}=\pm \frac{1}{\sqrt{\left| \mu \right| }}\lambda 
\label{lambda}
\end{equation}
with
\begin{equation}
\mu =\frac{\left(\pi \alpha \right)^{2}}{4g_{3}}
\left[\frac{\gamma_{p}^{2}}{\rho_L \left(c_{s}^{2}-\upsilon^{2}\right)}
-\frac{u_{\rho}^{2}-\upsilon ^{2}}{u_{\rho }\pi K_{\rho }}
+D_{t}\right]   \, .
\label{mu}
\end{equation}
Notice that (\ref{sinegordon}) is a sine-Gordon equation 
that has soliton solutions \cite{Chang}: 
\begin{eqnarray}
\Phi_{\rho}(x,t) &=& \pm \sqrt{2} \tan^{-1}\left( 
e^{\frac{1}{\sqrt{\left| \mu\right| }} \left(x-\upsilon t\right)}\right) 
,\text{ \ \ \ \ \ \ \ \ \ \ \ \ for }\mu >0  
\nonumber 
\\
&=&\pm \sqrt{2} \tan^{-1}\left( 
e^{\frac{1}{\sqrt{\left| \mu \right| }} \left( x-\upsilon t\right)}\right) 
+ \frac{\pi }{2\sqrt{2}},\text{ \ \ \ for } \mu <0.  
\label{solution for charge}
\end{eqnarray}
The electronic density that is related to $\Phi_{\rho}$
by equation (\ref{charge density operator}) and is localized in space
as in the polaron case (\ref{band insulator solution for electron})
with a width of order $\xi \approx 1/\sqrt{|\mu|}$.
In the present case, however, the soliton properties are heavily dressed
by the lattice and pseudo-spin. Therefore we will term
this excitation a {\it polaronic soliton}.

\section{The polaronic soliton at the Luther-Emery line}
\label{lutheremery}

In the previous section we described the physics of the polaronic soliton
formation in terms of the bosonized theory. Since the connection
between the electrons and boson fields is not linear, it is not
straightforward to interpret the formation of solitons in the bosonic
language as formation of solitons in the fermionic language. Moreover,
it is not clear what is the dressing process that changes an electronic
soliton into a polaronic soliton. 
To make the connection more explicit we are going to study the
problem of electron-phonon coupling along the so-called Luther-Emery
line. Along this line the bosonic degrees of freedom can be re-fermionized. 
To do so we re-introduce the backscattering
terms into the Hamiltonian and the parameters in the Hamiltonian
are chosen so that \cite{Emery,Solyom,Luther-Emery}:
\[
\frac{g_{1}-2g_{2}}{\pi \upsilon _{F}+g_{4}}=\frac{6}{5} \, ,
\]
where 
\[
\frac{g_{2}}{a}=\frac{g_{4}}{a}=U+2V \, .
\]
This is equivalent to choosing $K_{\rho}=1/2$. 
The charge part of the problem can be exactly diagonalized 
via an unitary transformation and the
final Hamiltonian reads \cite{Emery}:
\[
\bar{H}_{LE}=\sum_{k}E_{k}\left( \bar{C}_{1\rho k}^{\dagger }\bar{C}_{1\rho
k}-\bar{C}_{2\rho k}^{\dagger }\bar{C}_{2\rho k}\right) \, ,
\]
where $\bar{C}_{1\rho k}$ ($\bar{C}_{2\rho k}^{\dagger }$) is the
transformed annihilation (creation) operator for momentum $k$ for left 
(right) moving electrons. The energy spectrum is given by 
\begin{eqnarray}
E_{q}=\pm \frac{4}{5\pi }\left[ \left( \pi \upsilon _{F}+g_{4}\right)
^{2}q^{2}+\left( \frac{g_{3}}{2a}\right) ^{2}\right] ^{\frac{1}{2}} \, .
\label{mott dispersion}
\end{eqnarray}
Obviously the spectrum has a gap of magnitude $\Delta_M =\frac{4g_{3}}{5\pi a}$. This is the Mott gap. Notice that because $K_{\rho}=1/2$ this value of the
gap is in agreement with the RG result in (\ref{delta}).

We now add electron-phonon interaction 
(\ref{band insulator electron-phonon interaction}) into the Hamiltonian as 
\begin{eqnarray}
H_{LE-ph} = 
\sum_{k}E_{k}\left( \bar{C}_{1\rho k}^{\dagger }\bar{C}_{1\rho k}-
\bar{C}_{2\rho k}^{\dagger } \bar{C}_{2\rho k}\right) 
-\frac{\pi \gamma _{p}}{2} \sum_{p}
\sqrt{\frac{\hbar }{2M\omega \left( p\right) }} \, p \, \rho\left(p\right)
\left(a_{p}+a_{-p}^{\dagger }\right) \, ,
\end{eqnarray}
where $a_{p}^{\dagger }$(or $a_{p}$) is the phonon creation(or annihilation)
operator, $M$ is the mass of a unit cell, and
\begin{eqnarray}
\rho(p) = \sum_k \left(\bar{C}_{1\rho k+p}^{\dagger } \bar{C}_{1\rho k}+
\bar{C}_{2\rho k+p}^{\dagger } \bar{C}_{2\rho k}\right)
\end{eqnarray}
is the transformed fermionic density. Notice that the unitary tranformation
that diagonalizes the problem does not modify the form of the density operator.
What this result shows is that at the Luther-Emery 
problem becomes identical to the polaron problem discussed in Section 
\ref{band}. Since the operators $\bar{C}$ produce the solitonic excitations
in the electronic system it becomes clear that the polaronic soliton is
formed in two steps: (1) the electron-electron interaction generates
collective non-linear modes; (2) these non-linear modes are dressed by
the lattice fluctuations.

Let us make this description more concrete and assume that a soliton is
created in the upper band of the Mott insulator. 
We can expand (\ref{mott dispersion}) for small $q$ as:
\begin{eqnarray}
E^0_q \approx \frac{\Delta_M}{2} + \frac{q^2}{2 m^*} \, \, ,
\label{eok}
\end{eqnarray}
where 
\begin{eqnarray}
m^* = \left(\frac{5 \pi}{4}\right)^2 \frac{\Delta_M}{2 (\pi v_F + g_4)^2}
\label{mstar}
\end{eqnarray}
is the effective mass of the carrier.
Let us assume that 
the electron-phonon coupling is weak and compute the change in
energy of the ground state due to this coupling. It is a trivial exercise
in second order perturbation theory to show 
that for a state with momentum $q$ the energy changes by:
\begin{eqnarray}
\delta E_{q} &=& -\frac{\pi ^{2}\gamma _{p}^{2}}{2} \sum_{p}
\frac{\hbar p^{2}}{2 M L \omega\left(p\right)}
\frac{1}{E_{q+p}-E_{q}+\hbar \omega\left(p\right)}
\nonumber
\\
&=& - \frac{\pi m^*\gamma _{p}^{2}}{4 M \hbar c_{s}}
\ln \left\{\left[\frac{\hbar }{2\alpha 
\left(m^* c_{s}+\hbar q\right)}+1\right] 
\left[\frac{\hbar }{2\alpha \left( m^* c_{s}-\hbar q\right) }
+1\right]\right\} \, ,
\end{eqnarray}
where the integral was cut-off at $\Lambda \approx 1/\alpha$. 
For $\hbar q\ll m^*c_{s}$, this simplifies to
\begin{equation}
\delta E_{q}=-\frac{\pi m^*\gamma _{p}^{2}}{2M \hbar c_{s}}
\ln\left(\frac{\hbar}{2\alpha m^* c_{s}}+1\right) 
-\frac{\pi \gamma_p^{2}
\left(\hbar +4\alpha m^* c_{s}\right)}{4 \alpha M m^* c_{s}^{3}
\left(\hbar+2\alpha m^*c_{s}\right)^{2}}\hbar^{2} q^{2} \, .
\label{second order}
\end{equation}
By comparing (\ref{second order}) with (\ref{eok}) we conclude that
the energy of the carriers can be written as:
\begin{eqnarray}
E_q = \frac{\Delta_R}{2} + \frac{\hbar^2 q^2}{2 m_R} \, .
\end{eqnarray}
The Mott gap, $\Delta_R$, and the effective mass of the carriers,
$m_R$, have been 
renormalized by the polaronic effect and are given by:
\begin{eqnarray}
\Delta_{R} &=& \Delta_M - \frac{\pi m^* \gamma _{p}^{2}}{2M\hbar c_{s}}
\ln \left(\frac{\hbar}{2\alpha m^* c_{s}}+1\right)   
\nonumber 
\\
m_R &=& m^* \left[ 1+\frac{\pi \hbar \gamma_{p}^{2} g_{3}^{2}}{
8 m^* M c_{s}^{3}a^{2}\left( \pi \upsilon_{F}+g_{4}\right)^{2}}
\frac{\hbar \left(\hbar +4\alpha m^* c_{s}\right)}{
\left( \hbar +2\alpha m_{e}c_{s}\right) ^{2}}\right] \, .
\end{eqnarray}
and we conclude that the Mott gap is substantially reduced and the mass of
the carriers increased because of the polaronic effect.

\section{Conclusions}
\label{conclusions}

In this paper we have studied the problem of charge transport via
non-linear excitation in a band and Mott insulators. The insulating
behavior in band insulator is due to the periodicity of the lattice
and the electron-ion interaction opens a gap in the charge
spectrum. Mott insulators, on the other hand, are dominated by
electron-electron interactions and the gap is produced by correlation
effects when the electronic density is commensurate with the lattice. 
We argue that in band insulators the main propagating excitation is a polaron
that is described by an electron dressed by the polarization cloud.
In the Mott insulator, because of the strong electron-electron
interactions, the main non-linear excitation is the polaronic soliton,
a non-linear excitation of the Luttinger liquid that is dressed by
a polarization cloud.

We have shown that in a band insulator the doped electron (or hole)
is strongly dressed by the internal degrees of freedom in the system 
that we described in terms of phonons and pseudo-spins. We show
that the polaron can propagate freely: the lattice and the two-level
systems follow its motion in the system. 
Moreover, we have shown that the pseudo-spins
lead to an extra stabilization, relative to the phonon problem, 
by reducing its energy and size. 

In the case of the Mott insulator we have shown via a bosonization
calculation that the non-linear excitations can be thought of as
a soliton comprising an electron dressed by the phonons and
pseudospins. We term this new excitation the polaronic soliton. 
Many properties of these excitation are similar to the polaron
problem. Namely, the polaronic solitons are very stable and their
energy is reduced relative to the problem without phonons and
pseudo-spins. 

The great qualitative difference between a Mott insulator and a band
insulator is in the spin spectrum. Spin excitations in
a Mott insulator are
decoupled from the charge excitations (spin-charge separation) 
and remain gapless even when the Mott
gap opens. Thus Mott insulators have strong antiferromagnetic
correlations. This should be contrasted with the case of a band
insulator that has both a charge and a spin gap: in order to
excite an electron from the valence band to the conduction band
we have to pay an energy cost to unpair two electrons with opposite
spins in the valence band. In order to illustrate this point let us
consider a 1D system that is at half-filling with a weak periodic 
potential, $U(x)$, added such that this potential doubles the
unit cell. Because the system is half-filled, a potential that doubles
the unit cell has to open a gap at the Fermi energy. (This mechanism
is equivalent to the Peierls mechanism due to lattice distortions 
\cite{Heeger,alan}). Let us consider this process of opening a gap using
the bosonization technique. We consider the problem before the gap
opens. We can bosonize the system as explained in
Section \ref{introduction}. The coupling between the weak periodic
potential and the electrons can be written as:
\begin{eqnarray}
H_U &=& \int dx \sum_{\sigma} U(x) \psi^{\dag}_{\sigma}(x) \psi_{\sigma}(x) \, ,
\end{eqnarray}
where, because of the periodicity,
\begin{eqnarray}
U(x) = U_0 \cos(\pi x/a) \, .
\end{eqnarray}
We now expand the electron operators as in (\ref{slowf}) 
in terms of right and left moving electrons 
and find that in $H_U$ there are terms that oscillate with
$\cos(\pi x/a)$ and $\cos(\pi x/a \pm 2 k_F x)$. Since $k_F = \pi/(2 a)$ 
the terms with wave-vector $\pi/a \pm 2 k_F$ do not oscillate. All
the other terms vary very rapidly in space and can be disregarded. In summary,
we can rewrite $H_U$ as
\begin{eqnarray}
H_V &=& U_0 \int dx \sum_{\sigma} \psi^{\dag}_{R,\sigma}(x) \psi_{L,\sigma}(x)
+ h.c.
\nonumber
\\
&=& \frac{U_0}{\pi a} \int dx \sum_{\sigma} \cos(2 \sqrt{\pi} \phi_{\sigma}(x))
\nonumber
\\
&=& \frac{2 U_0}{\pi a} \int dx \cos(\sqrt{2 \pi} \Phi_{\rho}(x)) 
\cos(\sqrt{2 \pi} \Phi_{\sigma}(x)) \, ,
\label{hv}
\end{eqnarray}
where we have used the bosonization rules (\ref{boso1}) and 
(\ref{bosonization}). Notice that (\ref{hv}) describes the backscattering
of the electrons by the periodic potential of the lattice. In fact the
condition that $\pi/a = 2 k_F$ is nothing but the Bragg condition for
this particular problem. We have thus learned in the bosonized language
a very simple fact about electron scattering in periodic potentials.
Note, however, that in the bosonic language the extra periodic potential
is a highly non-linear operator in terms of the bosons. In practice,
this term is not easy to treat exactly. Let us consider, however,
a perturbative RG calculation like the one explained in Section 
\ref{introduction}. It is easy to show that:
\begin{eqnarray}
\frac{\partial U_0}{\partial \ell} = \left(2-\frac{K_{\rho}}{2}-
\frac{K_{\sigma}}{2} \right) U_0 \, ,
\label{rgv0}
\end{eqnarray}
which demonstrates that the periodic potential is a relevant perturbation of
the Luttinger liquid when $K_{\rho}+K_{\sigma} < 4$. Notice that
when this happens (as we explained in Section \ref{introduction})
$H_U$ becomes relevant and the fields $\Phi_{\rho}$ and
$\Phi_{\sigma}$ are pinned at the minimum of the potential energy.
In this case, both a charge gap and a spin gap open in the spectrum!
The size of such gaps, as in (\ref{delta}), can be obtained from
the RG (\ref{rgv0}) and read:
\begin{eqnarray}
\Delta_U = W \left(\frac{U_0}{W}\right)^{\frac{1}{2-\frac{K_{\rho}}{2}-
\frac{K_{\sigma}}{2}}} \, .
\label{deltau}
\end{eqnarray}
This result shows that a gap in a correlated band insulator depends on
the strength of the electron-electron interactions. In particular,
in the non-interacting limit when $K_{\rho}=K_{\sigma}=1$ we
see that $\Delta_M^0 = U_0$, as expected from elementary solid state physics.
It is clear that in the case of a band
insulator the spin gap has nothing to do with superconducting fluctuations
but rather with the fact that the band is full and therefore 
that the electronic shells are closed.

The RG flow for the periodic potential (\ref{rgv0}) 
should be contrasted with
the RG flow for the Umklapp term, Eq.(\ref{rg2}), 
which opens the Mott gap, and
is relevant for $K_{\rho}<1$. Notice that there is a
competition between the Umklapp term and the periodic potential:
for $0 < K_{\rho} <K_{\sigma}/3$ the Umklapp term diverges faster than the
periodic potential while for $K_{\sigma}/3<K_{\rho}<4-K_{\sigma}$
the electron-ion interaction has a stronger divergence. 
(For $K_{\rho}>4-K_{\sigma}$ both operators are irrelevant). 
Since $K_{\rho}$ decreases with the strength of
the electron-electron interaction it becomes obvious that for strong
electron-electron interaction the Umklapp term is more important
than the periodic potential and a Mott gap opens in the system.
The spin spectrum is gapless while the charge spectrum has
a gap. On the other hand, if the electron-electron interactions are weaker,
the periodic potential dominates and a band gap opens in both
the spin and the charge spectrum. It is clear that we can
differentiate the band gap and the Mott gap by examining the spin
spectrum. The critical value of interactions, $U_c$ and $V_c$ in
(\ref{Extended Hubbard Model}), above which
a Mott gap appears in a band insulator occurs when $K_{\rho}=
K_{\sigma}/3$. Using (\ref{u & K & g1 & g3}) we find
\begin{eqnarray}
U_c - \frac{2}{5} V_c = \frac{4 \pi v_F}{5 a} \, ,
\end{eqnarray}
which gives the critical line in the $U \times V$ plane above which
one finds a Mott insulator and below which we have a band insulator.

The applications of the formalist developed here to systems like DNA
depends very much on the actual values of $U$ and $V$. 
If DNA is better described by a band insulator then the
formalism developed in section \ref{band} is more appropriate
and the Davydov's equations (\ref{davydov}) should apply. In this
case the elementary excitations should be polarons. However,
if the electron-electron interactions are strong, as we conjecture, 
then the polaronic soliton of the Mott insulator, 
as described in section \ref{mott},
is the best description. While numerical simulations 
can provide useful information on order of magnitude of the electronic
parameters, only well-controlled experiments in periodic DNA
sequences can actually give the final word on the insulating nature of
DNA. As we discussed previously, we believe that the correct way
to find out about the insulating nature of DNA is by 
the study of the spin excitation spectrum and not by conductivity
measurements that are insensitive to the spin degrees of freedom.
While there are indications of antiferromagnetic fluctuations in
DNA that would be consistent with the Mott insulator picture \cite{barton}
the experimental picture is far from complete and many more experiments
are needed. 

In summary, we have shown that non-linear excitations in band and
Mott insulators are very similar and have solitonic character. We show
that the band gap is strongly renormalized by electron-electron interactions
in 1D and that a 1D system can be driven from a band insulator type
behavior to a Mott insulator behavior as a function of interactions.
The correct way to distinguish between these two types of insulator it is
not by studying the charge degrees of freedom, that propagate in the form
of non-linear waves, but to the spin degrees of freedom that remain 
gapless in the Mott case and are gapped in a band insulator. We believe
that the physics described here can be applied to strongly correlated systems
like DNA. 

We thank illuminating discussions with J.~Barton, W.~Beyermann, D.~Cox,
G.~Gruner, and S.~Kivelson. We thank Allen G. Hunt for organizing
Michael Pollak's Festschrift. We especially thank M.~Pollak for
estimulating our interest in this problem. We acknowledge 
partial support provided by the Collaborative University of California - Los 
Alamos (CULAR) research grant under the auspices
of the US Department of Energy.

\end{document}